\documentclass[fleqn,usenatbib]{mnras}

\usepackage[T1]{fontenc}
\usepackage{ae,aecompl}

\usepackage{graphicx}	% Including figure files
\usepackage{amsmath}	% Advanced maths commands
\usepackage{amssymb}	% Extra maths symbols
\usepackage{float}
\usepackage[justification=centering]{caption} % Caption
\usepackage{multirow} % For table 1
\usepackage{booktabs} % For table 1

\usepackage{newtxtext,newtxmath}

\title[]{Galaxy And Mass Assembly: Galaxy Zoo spiral arms and star formation rates}

\author[R. Porter-Temple et al.]{R. Porter-Temple$^{1}$, 
B. W. Holwerda$^{1}$\thanks{Contact e-mail: \href{mailto:benne.holwerda@louisville.edu}{benne.holwerda@louisville.edu}},
A. M. Hopkins$^{2}$, 
L. E. Porter$^{1}$, 
C. Henry$^{1}$,
\newauthor T. Geron$^{3}$,
B. Simmons $^{4}$,
K. Masters$^{5}$, and
S. Kruk$^{6}$
\\
$^{1}$ University of Louisville, Department of Physics and Astronomy, 102 Natural Science Building, 40292 KY Louisville, USA.\\
$^{2}$ Australian Astronomical Optics, Macquarie University, 105 Delhi Rd, North Ryde, NSW 2113, Australia\\
$^{3}$ Department of Physics, University of Oxford, Denys Wilkinson Building, Keble Road, Oxford OX1 3RH, UK\\
$^{4}$ Physics, Lancaster University, Lancaster LA1 4YB, UK \\
$^{5}$ Departments of Physics and Astronomy, Haverford College, 370 Lancaster Ave, Haverford, PA 19041, USA \\
$^{6}$ Max-Planck-Institut f\"ur extraterrestrische Physik (MPE), Giessenbachstrasse 1, D-85748 Garching bei M\"unchen, Germany\\
}

\date{Accepted XXX. Received YYY; in original form ZZZ}

\pubyear{2022}

\begin{document}
\label{firstpage}
\pagerange{\pageref{firstpage}--\pageref{lastpage}}
\maketitle

\begin{abstract} \label{abstract}

Understanding the effect spiral structure has on star formation properties of galaxies is important to completing our picture of spiral structure evolution.  Previous studies have investigated connections between spiral arm properties with star formation, but the effect that the number of spiral arms has on this process is unclear. Here we use the Galaxy and Mass Assembly (GAMA) survey paired with the citizen science visual classifications from the Galaxy Zoo project to explore galaxies' spiral arm number and how it connects to the star formation process. We use the votes from the GAMA-KiDS GalaxyZoo classification to investigate the link between spiral arm number with stellar mass, star formation rate, and specific star formation rate. We find that galaxies with fewer spiral arms have lower stellar masses and higher sSFRs, while those with more spiral arms tend toward higher stellar masses and lower sSFRs, and conclude that galaxies are less efficient at forming stars if they have more spiral arms. We note how previous studies' findings may indicate a cause for this connection in spiral arm strength or opacity.
\end{abstract}

\begin{keywords}
galaxies: spiral
galaxies: star formation 
galaxies: statistics 
galaxies: stellar content 
galaxies: structure
\end{keywords}

%%%%%%%%%%%%%%%%%%%%%%%%%%%%%%%%%%%%%%%%%%%%%%%%%%

%%%%%%%%%%%%%%%%% BODY OF PAPER %%%%%%%%%%%%%%%%%%

\section{Introduction} \label{introduction}

% Galaxy zoo flowchart
\begin{center}
\begin{figure*}
        \includegraphics[width=0.65\textwidth]{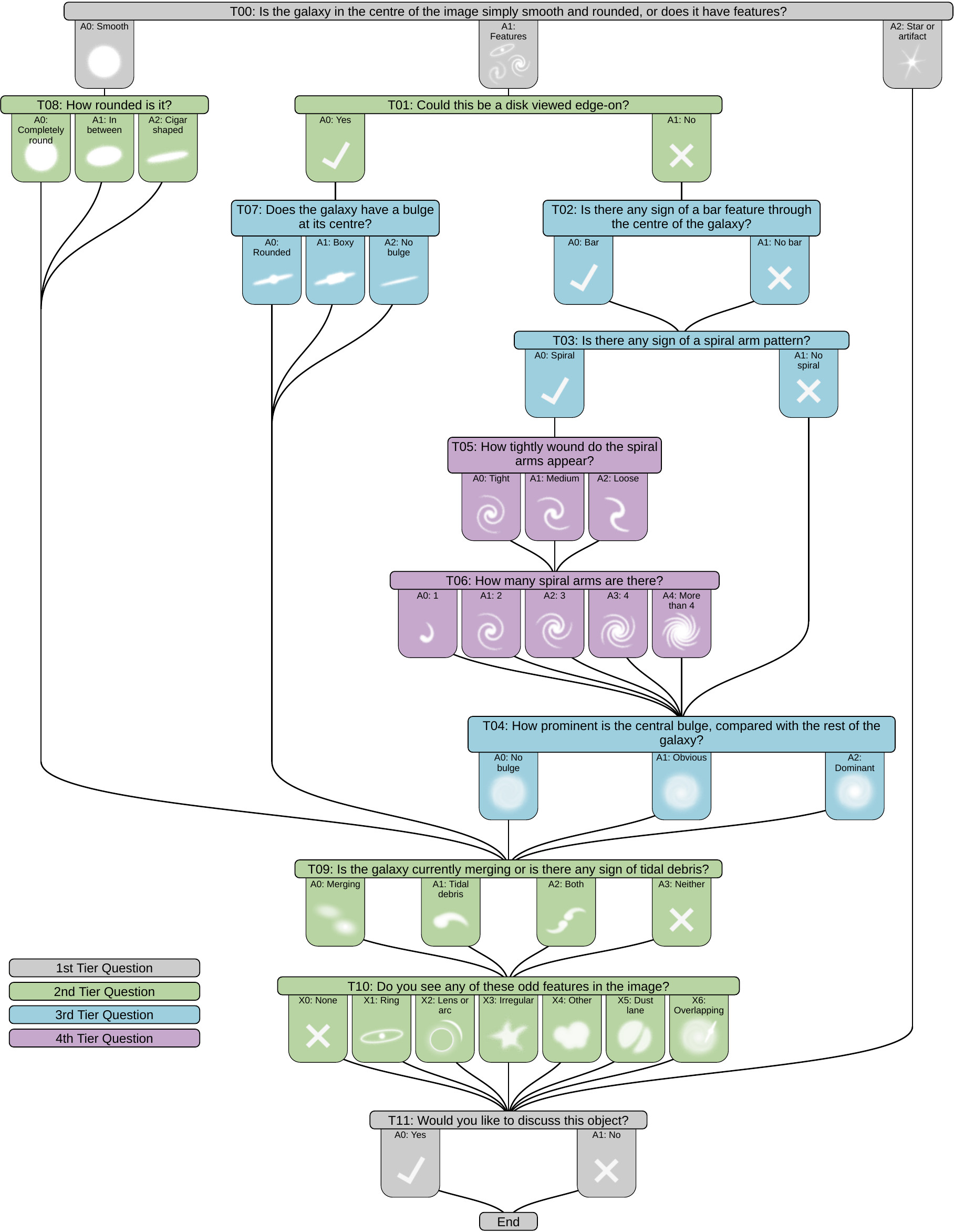}
        \caption{Galaxy Zoo 4 GAMA-KiDS decision tree. The decision tree can be viewed at \url{https://data.galaxyzoo.org/gz_trees/gz_trees.html} under the 'GZ GAMA-KiDS' section.}
        \label{fig:GZ4}
\end{figure*}   
\end{center}

%intro
Though they are visually distinctive, the exact properties that affect the formation of arms in spiral galaxies are not yet fully explored. 
The potential and observed links between spiral galaxy structure and pitch angle, arm strength \citep{Seigar98,Yu20,Kendall14} are now being examined with increasing samples sizes and sophistication in analysis \citep{Hart17a,Yu18a,Lingard21}. The motivation for increased interest in spiral morphology in representative samples is to constrain the dominant formation mechanism of spiral arm structure \citep{Masters21}. For example, \cite{Pringle19a} find a constant distribution of pitch angles of arms, consistent with the density wave theory origin of spiral structure. \cite{Hart18} found 40 per cent of arm formation in massive spirals can be by ``swing amplification"; the number of arms is consistent with the prediction from this mechanism with the remainder originating from other mechanisms. % weird that 60% is "the remainder"

\cite{Diaz-Garcia19c} do not find observational evidence that spiral arms are driven by stellar bars \citep[as do][]{Hart17a} or through ``manifolds", pathways of infalling material, which would show as a dependence of arm strength and pitch angle \citep{Athanassoula09}. They found that bar and arm strength are correlated, while bar strength and pitch angle are not. In multi-wavelength data, \cite{Yu18b} found younger stars to reside in tighter arms and \cite{Miller19a} found that these stars then trailed out from the arms. 

\cite{Seigar98} found no correlation between pitch angle and Hubble type, which was reiterated by \cite{Kendall14, Yu18}; however, they note that not finding a correlation is unsurprising given the small range of pitch angles they examined. \cite{Yu20} later found a loose correlation between pitch angle and spiral arm strength, with an overall tendency for pitch angle to decrease with weaker arm strength, while \cite{Savchenko20} find no strong difference (except for number of arms) between grand design, multi-armed, and flocculent spirals in pitch angle, arm width or strength. The link between arm strength, pitch angle and formation mechanisms remains complex. 

Instead of focusing on the formation of spiral arm structure, one can examine the correlations with global properties of the galaxies such as star formation, stellar mass or specific star formation.
\cite{Hart17} investigated spiral structure using the Sloan Digital Sky Survey (SDSS) main galaxy sample, with morphological data from the public release of Galaxy Zoo 2 \citep{Willett13}, stellar mass from \cite{Chang15}, and star formation from GALEX fluxes \citep{Martin05}. Using these data, they determined no significant dependence of spiral arm number on specific star formation rate (sSFR). 

%\cite{Hart17} investigated spiral structure, and determined no significant dependence of specific star formation rate (SSFR) on spiral arm number, based on the Sloan Digital Sky Survey (SDSS) main galaxy sample with morphological data from the public release of Galaxy Zoo 2 \citep{Willett13} and stellar mass from \cite{Chang15} and star formation from GALEX fluxes \cite{Martin05}.

In this paper, we explore the connection of spiral arm number with stellar mass, star formation rate (SFR), and specific star formation rate (sSFR), using similar methods to \cite{Hart17}. We make use of the improved star formation and stellar mass estimates by the Galaxy And Mass Assembly \citep[GAMA,][]{Driver09,Liske15} survey using self-consistent {\sc magphys} SED fits to the full uv to sub-mm SED \citep{Driver16c,Wright16} and GalaxyZoo voting base on deeper and higher-resolution KiDS data \cite[][Kelvin et al. \textit{in prep.}]{Holwerda19b}. With this improved quality data, we investigate the trends with spiral arm numbers that the results from \cite{Hart17} suggested. We compare spiral arm number subsamples of stellar mass, SFR, and sSFR to the whole set of galaxies to determine any notable differences. The sets defined by a spiral number (m = 1, 2, 3, 4, 5+) are from the visual classification from the GAMA-KiDS Galaxy Zoo project, detailed in Section \ref{galaxy zoo}. This paper is organized as follows: section \ref{data} describes the data used in the paper and how subsamples are defined, section \ref{results} presents the results for star formation, stellar mass, and specific star formation as a function of the number of spiral arms, 
section \ref{discussion} discusses these results and section \ref{conclusions} lists our conclusions.

%% tables %%
\begin{table*}
    \large
    \centering
    \renewcommand{\arraystretch}{1.2} 
    %\scalebox{1.1}{
    \begin{tabular}{llcccccccccccc} \toprule 
        & & \multicolumn{2}{c}{2-Sample K-S Test} & \multicolumn{2}{c}{k-Sample A-D Test} & \multicolumn{7}{c}{k-Sample A-D Test Critical Values} \\ \cmidrule(lr){3-4} \cmidrule(lr){5-6} \cmidrule(lr){7-13}
        & &  Statistic & Significance & Statistic & Significance & 25\% & 10\% & 5\% & 2.5\% & 1\% & 0.5\% & 0.1\%\\\midrule
        \multirow{5}{*}{ \rotatebox{90}{Stellar Mass} } 
        & m=1 & 0.138 & \textbf{0.038} & 3.977 & 0.008 & 0.33 & 1.23 & 1.96 & 2.72 & \textbf{3.75} & 4.59 & 6.55\\
        & m=2 & 0.034 & 0.203 & 0.763 & 0.159 & \textbf{0.33} & 1.23 & 1.96 & 2.72 & 3.75 & 4.59 & 6.55\\
        & m=3 & 0.152 & \textbf{0.001} & 6.642 & 0.001 & 0.33 & 1.23 & 1.96 & 2.72 & 3.75 & 4.59 & \textbf{6.55}\\
        & m=4 & 0.252 & 0.081 & -0.071 & 0.250 & \textbf{0.33} & 1.23 & 1.96 & 2.72 & 3.75 & 4.59 & 6.55\\
        & m=5+ & 0.264 & \textbf{0.000} & 10.216 & 0.001 & 0.33 & 1.23 & 1.96 & 2.72 & 3.75 & 4.59 & \textbf{6.55}\\ \midrule
        \multirow{5}{*}{ \rotatebox{90}{SFR} } 
        & m=1 & 0.099 & 0.256 & 0.278 & 0.250 & \textbf{0.33} & 1.23 & 1.96 & 2.72 & 3.75 & 4.59 & 6.55\\
        & m=2 & 0.110 & \textbf{0.000} & 36.514 & 0.001 & 0.33 & 1.23 & 1.96 & 2.72 & 3.75 & 4.59 & \textbf{6.55}\\
        & m=3 & 0.281 & \textbf{0.000} & 39.523 & 0.001 & 0.33 & 1.23 & 1.96 & 2.72 & 3.75 & 4.59 & \textbf{6.55}\\
        & m=4 & 0.291 & \textbf{0.028} & 4.342 & 0.006 & 0.33 & 1.23 & 1.96 & 2.72 & \textbf{3.75} & 4.59 & 6.55\\
        & m=5+ & 0.217 & \textbf{0.004} & 7.721 & 0.001 & 0.33 & 1.23 & 1.96 & 2.72 & 3.75 & 4.59 & \textbf{6.55}\\ \midrule
        \multirow{5}{*}{ \rotatebox{90}{sSFR} } 
        & m=1 & 0.234 & \textbf{0.000} & 9.270 & 0.001 & 0.33 & 1.23 & 1.96 & 2.72 & 3.75 & 4.59 & \textbf{6.55}\\
        & m=2 & 0.072 & \textbf{0.000} & 19.745 & 0.001 & 0.33 & 1.23 & 1.96 & 2.72 & 3.75 & 4.59 & \textbf{6.55}\\
        & m=3 & 0.187 & \textbf{0.000} & 10.309 & 0.001 & 0.33 & 1.23 & 1.96 & 2.72 & 3.75 & 4.59 & \textbf{6.55}\\
        & m=4 & 0.261 & 0.064 & 0.901 & 0.139 & \textbf{0.33} & 1.23 & 1.96 & 2.72 & 3.75 & 4.59 & 6.55\\
        & m=5+ & 0.143 & 0.126 & 1.976 & 0.050 & 0.33 & 1.23 & \textbf{1.96} & 2.72 & 3.75 & 4.59 & 6.55\\\bottomrule
\end{tabular}
\caption{\label{tab:statstable} Spiral arm number (m), Kolmogorov-Smirnov test statistic and significance for stellar mass, star formation rate, and specific star formation rate are shown under the header 2-sample K-S Test. Bold values for the K-S test significance are the statistically significant values discussed in section \ref{discussion}. The Anderson Darling test statistic and estimated significance level for Stellar Mass, SFR, and sSFR are shown under the header k-sample A-D Test. The critical values for different levels of significance are listed, with the critical value that each subsample meets in bold. The Anderson Darling test significance estimates are floored at 0.1\% and capped at 25\%.}
    %scaleboxend }
\end{table*}
%% tables end %%

% \textbf{Additional Papers
% }
%  \cite{Hart17b} investigate further into the link between pitch spiral arm angle and bars. 

% \cite{Yu18} - pitch angle is smaller (more tightly wound) in massive galaxies and/or with prominent bulge. Link between pitch angle and rotation curve: steeply rising-tightly wound arms.
% \cite{Yu18a} - how to measure pitch angle from Fourier analysis of the images.
% \cite{Yu18b} - younger stars in tighter arms than older stars
% \cite{Savchenko20} - find no strong difference (except for  number of arms) between grand design, multi-armed, and flocculent spirals in pitch angle, arm width or strength.

% \cite{Pringle19a} find a constant distribution of pitch angle, a prediction of the density wave theory origin of spiral structure

% \cite{Miller19a} measures pitch angles across wavelength finding consistency with younger stars trailing the density wave.

% \cite{Lingard21}-  a hierarchical Bayesian approach to galaxy pitch angle determination, 

% \cite{Hart18} - 40 per cent of arm formation in massive spirals is driven by ``swing amplification" (a resonance effect; the number of arms is consistent with the prediction from this mechanism. The remainder must be triggered by other effects (tidal etc) 

% \cite{Diaz-Garcia19c} do not find observational evidence that spiral arms are driven by stellar bars or ``manifolds" (pathways of incoming material) which would show a dependence on arm strength and pitch angle \citep{Athanassoula09}. Bar and arm strength are correlated, bar strength and pitch angle are not.

\section{Data} \label{data}
    
The data used comes from the Galaxy and Mass Assembly (GAMA) survey \citep{Driver09, Liske15}. We use the GAMA DR3 \citep{Baldry18} and the Kilo Degree Survey \citep[KiDS,][]{de-Jong13,de-Jong15,de-Jong17,Kuijken19} imaging. Additionally, we use the MAGPHYS table described in the GAMA DR3. MAGPHYS computes the stellar mass and specific star formation rate used and is fully described in \cite{da-Cunha08}.

\subsection{GAMA} \label{gama}
GAMA is a combined spectroscopic and multi-wavelength imaging survey designed to study spatial structure in the nearby ($z < 0.25$) Universe on kpc to Mpc scales \citep[see][for an overview]{Driver09, Driver11}. The survey, after completion of phase 2 \citep{Liske15}, consists of three equatorial regions each spanning 5 deg in Dec and 12 deg in RA, centered in RA at approximately 9h (G09), 12h (G12) and 14.5h (G15) and two Southern fields, at 05h (G05) and 23h (G23). The three equatorial regions, amounting to a total sky area of 180 deg$^2$, were selected for this study. For the purpose of visual classification, 49,851 galaxies were selected from the equatorial fields with redshifts $z<0.15$ (see below).   The GAMA survey is $>$98\% redshift complete to r $<$ 19.8 mag in all three equatorial regions. We use the  {\sc magphys} SED fits data-products \citep{Driver18} from the third GAMA data-release \citep[DR3,][]{Baldry18}.

\subsection{KiDS} \label{kids}

The Kilo Degree Survey \citep[KiDS,][]{de-Jong13,de-Jong15,de-Jong17,Kuijken19} is an ongoing optical wide-field imaging survey with the OmegaCAM camera at the VLT Survey Telescope. It aims to image 1350 deg$^2$ in four filters (u g r i). The core science driver is mapping the large-scale matter distribution in the Universe, using weak lensing shear and photometric redshift measurements. Further science cases include galaxy evolution, Milky Way structure, detection of high-redshift clusters, and finding rare sources such as strong lenses and quasars.
KiDS image quality is typically 0\farcs6 resolution (for sdss-r) and depths of 23.5, 25, 25.2, 24.2 magnitude for i, r, g and u respectively. This imaging was the input for the GalaxyZoo citizen science classifications.

\subsection{Galaxy Zoo} \label{galaxy zoo}

Information on galaxy morphology is based on the GAMA-KiDS Galaxy Zoo classification \citep[][Kelvin et al., \textit{in prep.}]{Lintott08}. The GAMA-KiDS Galaxy Zoo project is described in Kelvin et al., \textit{in prep}. RGB cutouts were constructed from KiDS g-band and r-band imaging with the green channel as the mean of these. KiDS cutouts were introduced to the classification pool and mixed in with the ongoing classification efforts. 
For the Galaxy Zoo classification, 49,851 galaxies were selected from the equatorial fields with redshifts $z < 0.15$. The Galaxy Zoo provided a monumental effort with almost 2 million classifications received from over 20,000 unique users over the course of the first 12 months. 
This classification has been used by the GAMA team to identify dust lanes in edge-on galaxies \citep{Holwerda19},  searches for strong lensing galaxy pairs \citep{Knabel20}, and the morphology of green valley galaxies (Smith et al {\em in prep}.).
In this paper we use the visual classifications of spiral galaxies from the Galaxy Zoo project; the full decision tree for the GAMA-KiDS Galaxy Zoo project is shown in Figure \ref{fig:GZ4}.

% The GAMA-KiDS Galaxy Zoo project uses the decision tree in use for the latest (4th) iterations of the Zoo. KiDS cutouts were introduced to the classification pool and mixed in with the ongoing classification efforts. Scientific aims include correlating general morphology to the GAMA results using the full suite of multi-wavelength and spectral information and the identification of rare features (e.g. strong lensing arcs of galaxy occultation). 
% A full description of the GAMA-KiDS Galaxy Zoo effort can be found in Kelvin et al. {\em in preparation}.
% This classification has been used by the GAMA team to identify dust lanes in edge-on galaxies \cite{Holwerda19} and searches for strong lensing galaxy pairs \citep{Knabel20}.
%% MAGPHYS %%
\subsection{MAGPHYS SED} \label{magphys}
In addition to the GAMA-KiDS Galaxy Zoo classifications, we use the {\sc magphys} \citep{da-Cunha08}, spectral energy distribution fits to the GAMA multi-wavelength photometry \citep{Wright17}, presented in \cite{Driver18}. {\sc magphys} computes stellar mass,  star formation rate, and specific star formation rates which will serve as comparison data for the Galaxy Zoo arm classifications.

%% sample selection %%
\subsection{Sample Selection} \label{sample selection}
\begin{figure}
        \centering
        \includegraphics[scale=0.49]{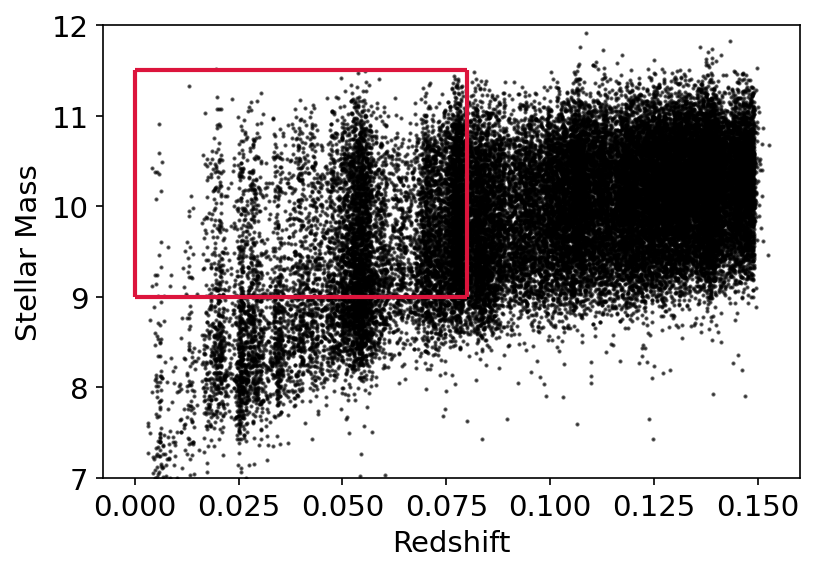}
        \caption{Stellar Mass vs. Redshift for the GAMA-KiDS Galaxy Zoo project data. The limited sample, which includes only those galaxies with $z < 0.08$ and $M_* > 10^9$, is indicated by the red box. Only galaxies with 30\% or more votes in favor of being a spiral galaxy are included.}
        \label{fig:redshiftLIM}
\end{figure}
To be included in the subset of the GAMA-KiDS Galaxy Zoo project used (herein after referred to as 'the limited sample'), a galaxy must meet three criteria. First, the galaxy must have a stellar mass \(M_* > 10^9\). Any galaxies below that limit are excluded. Second, the galaxy must have received at least 30\% of votes in favor of it being a spiral galaxy. This is represented by question T03 in the Galaxy Zoo decision tree shown in Figure \ref{fig:GZ4}. This avoids galaxies that were misclassified as spiral galaxies due to a low number of votes. Third, included galaxies must have a redshift less than 0.08, meaning any galaxies with $z \geq 0.08$ are not included in the limited sample. Doing this excludes those galaxies whose spiral arms are not correctly represented by Galaxy Zoo votes because of unclear imaging or lack of distinction between, for example, two-armed and four-armed spirals at $z\geq 0.08$. 

The limits on the limited sample from the full GAMA-KiDS Galaxy Zoo project is shown in Figure \ref{fig:redshiftLIM}.

%% subsamples %%
\subsection{Defining Subsamples} \label{subsample}

Each subsample of spiral galaxies is defined by their spiral arm number as voted by Galaxy Zoo participants. This is represented by question T06 in Figure \ref{fig:GZ4}, with answers A0, A1, ..., A4 being classified in this paper as m=1, m=2, ..., m=5+. 

In addition to fulfilling all the criteria described in section \ref{sample selection}, to fall into any given subsample m=x, a galaxy must meet two additional criteria. First, it must have received at least 50\% of votes in favor of having x spiral arms; that is, a galaxy is in the m=x subsample if the fraction of votes in favor of x arms is \( > 0.5\). The cutoff at 50\% means that the majority of votes dictates what subsample the galaxy falls into, so no galaxy falls into more than one subsample. Second, the galaxy must have less than 100\% of votes in favor of it having x spiral arms. This eliminates some galaxies that have a very low number of votes. So, a galaxy that with a fraction of votes $f_m$ in the range $(0.5 < f_m < 1)$ for answer A0 in Table \ref{fig:GZ4} would be included in the m=1 subsample, and similar for m=2, 3, 4, and 5+ spiral arms. 

%%%% RESULTS %%%%
\section{Results} \label{results}

The limited sample of galaxies is compared with each subsample as determined in section \ref{subsample}, with respect to stellar mass, SFR, and sSFR. The number of galaxies N given in each subsample is shown in Table \ref{tab:statstable}, along with the Kolmogorov–Smirnov test (K-S test) statistic and p-value. 

The K-S test statistic indicated how similar the subsample is to the parent sample, with smaller values being more similar and larger values being less similar, where a statistic of 0.0 indicates two identical distributions. The p-value associated with each K-S statistic dictates the significance in the K-S statistic, and we consider a p-value of .05 or lower to be significant.

% A-D Test
For an additional test of sample similarity, we perform the Anderson-Darling test on the above samples with the resulting statistic and p-values also listed in Table \ref{tab:statstable}. The critical values for each A-D test are returned for different levels of confidence and we bold the value that is exceeded by the A-D statistic in each case. The benefit of the A-D test over the K-S test is that it identifies confidence levels independently from the reported p-value. The A-D test is much more sensitive to the tails of any distribution and the K-S test is more dependent of the center of distribution. As our distributions are all non-Gaussian, this makes the A-D test better suited for the comparison.

% Differences in K-S and A-D results
Broadly the K-S and A-D tests agree on which populations differ but they disagree on the level of significance. For example Stellar Mass and one arm (m=1) or SFR (m=4), the A-D test assigns higher significance to the difference. We note that the K-S test reports a small, but low significance difference for the m=5+ sSFR distribution but the A-D identified a (just) significant result (5\% critical value exceeded, Table \ref{tab:statstable}).

%% results figures %%
    % stellar mass histograms
\begin{figure*}
    \centering
        \includegraphics[width=.32\textwidth]{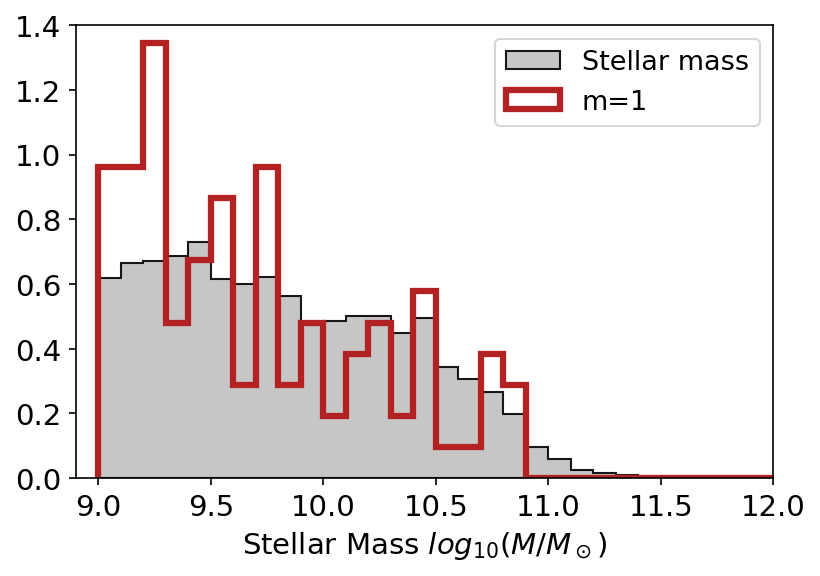}\quad
        \includegraphics[width=.32\textwidth]{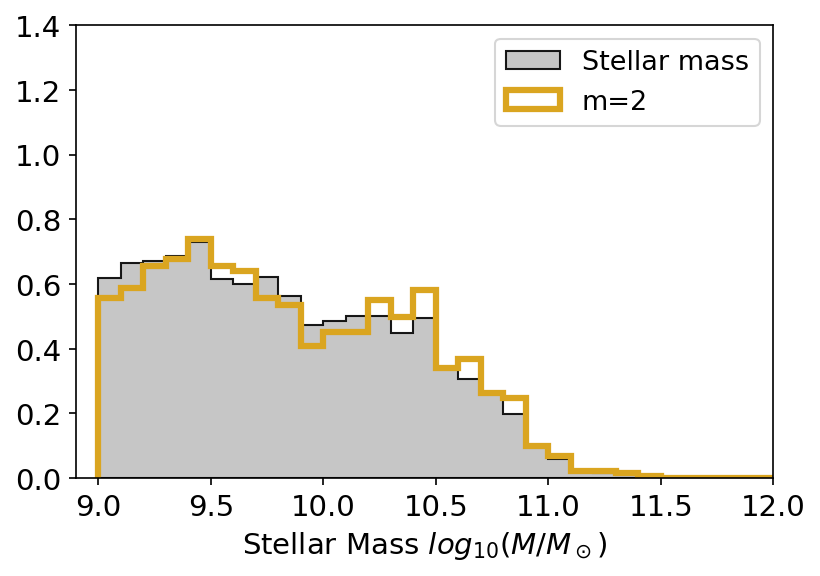}
    \smallskip
        \includegraphics[width=.32\textwidth]{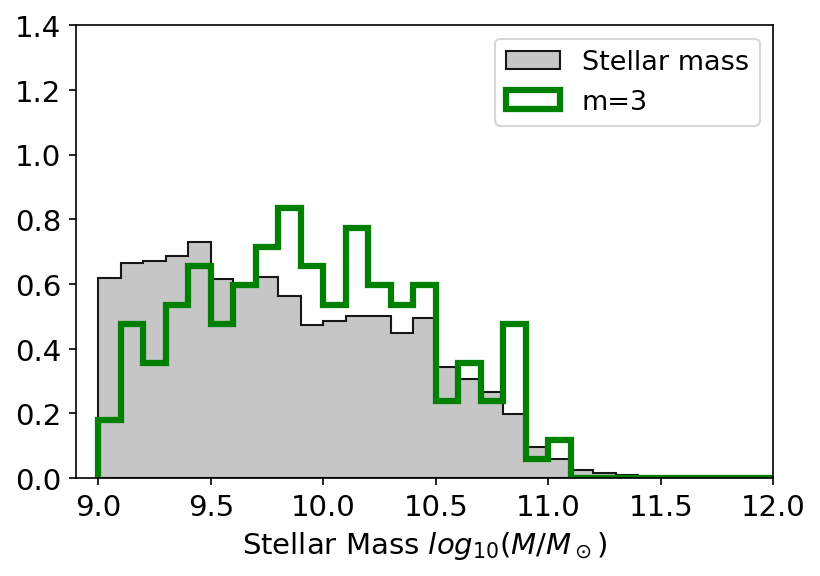}\quad
        \includegraphics[width=.32\textwidth]{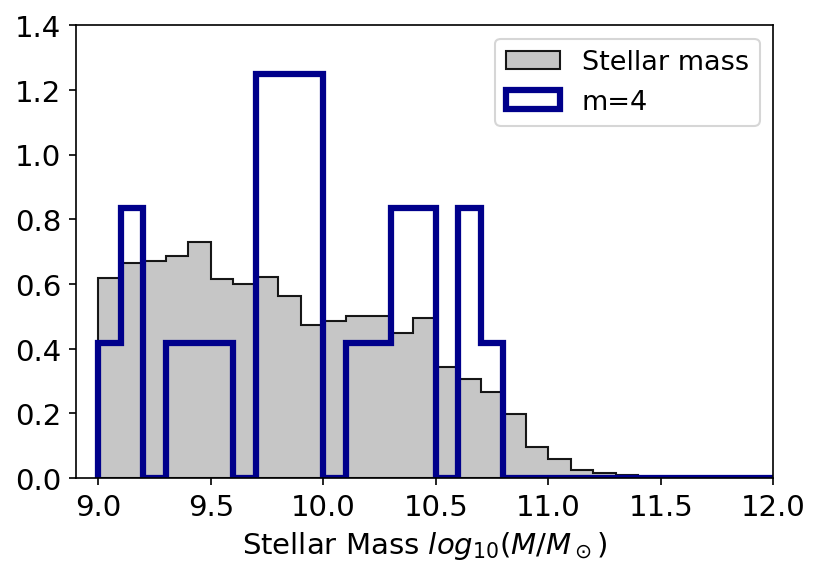}
    \smallskip
        \includegraphics[width=.32\textwidth]{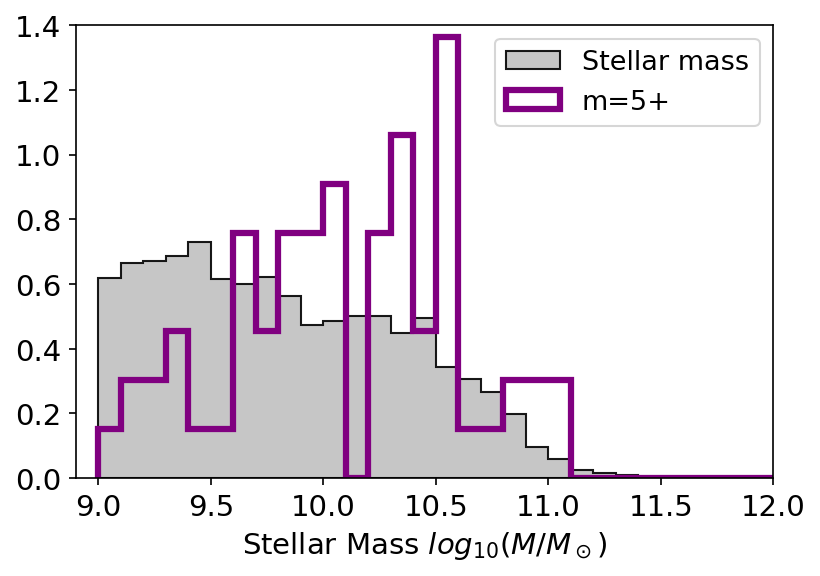}
    \caption{Stellar mass histograms for each of the subsamples selected from the limited GAMA-KiDS Galaxy Zoo sample. The gray filled histogram shows the distributions of the entire limited data set, while the colored outlines show the distribution for the individual spiral arm number subsamples.}
    
    \label{pics:stellarmassfig}
\end{figure*}

    % stellar mass summary
\begin{figure}
   \centering
    \includegraphics[width=0.49\textwidth]{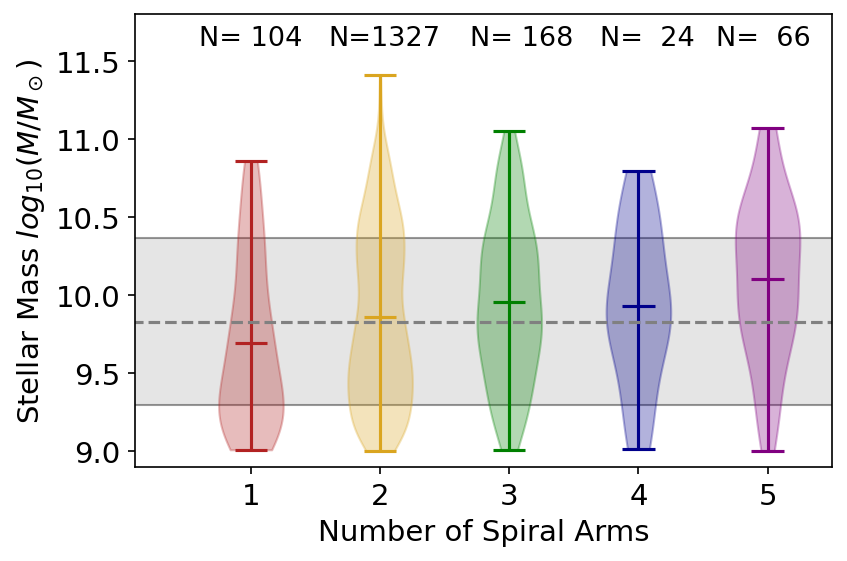}
    \caption{Stellar mass distribution densities for each of the spiral arm subsamples. The shaded gray region indicates $\pm$1 standard deviation of the whole sample. The dotted gray line indicates the mean for the whole sample. Each spiral arm distribution shows the high range, mean, and low range, indicated by horizontal dash marks. The number of galaxies in each subsample is shown above each distribution.}
    \label{fig:stmsummary}
\end{figure}

    % SFR histograms
\begin{figure*}
    \centering
        \includegraphics[width=.32\textwidth]{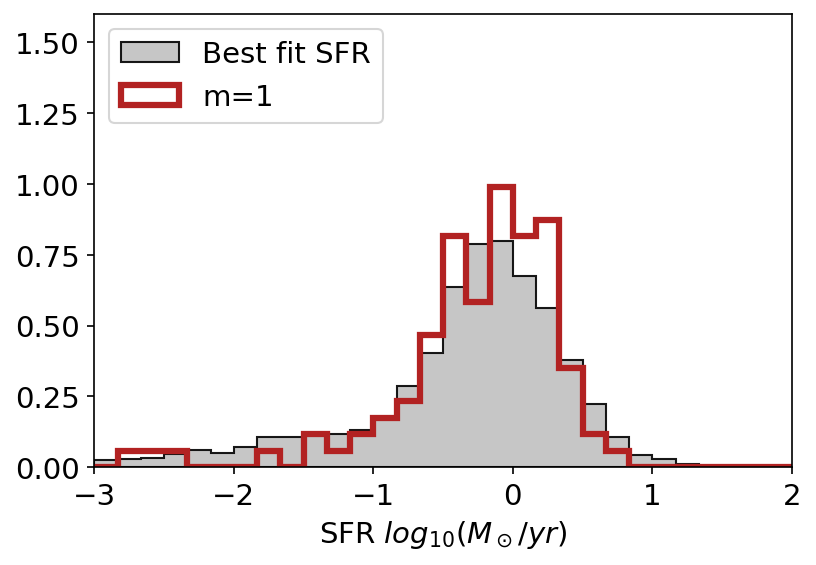}\quad
        \includegraphics[width=.32\textwidth]{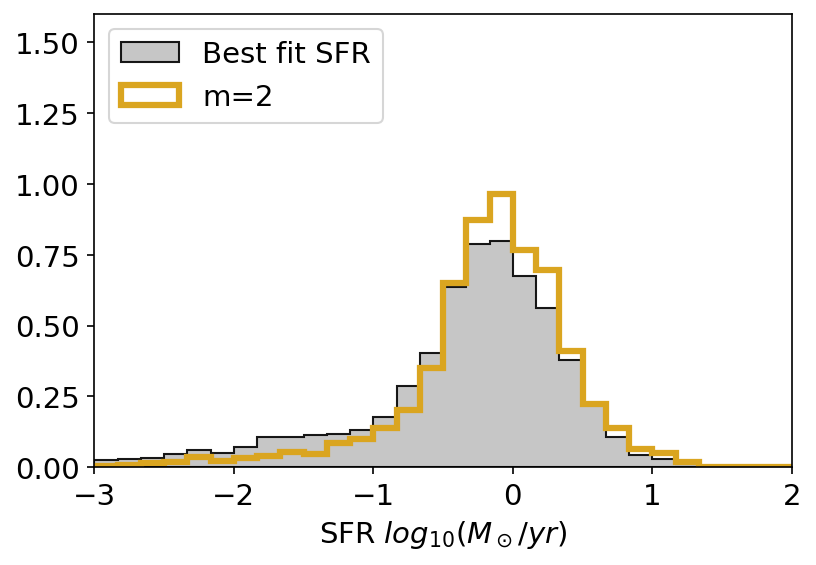}
    \smallskip
        \includegraphics[width=.32\textwidth]{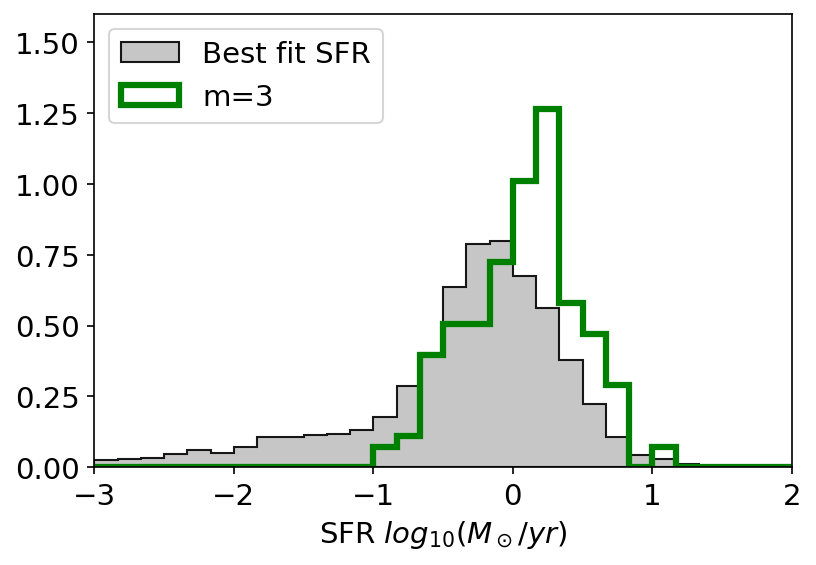}\quad
        \includegraphics[width=.32\textwidth]{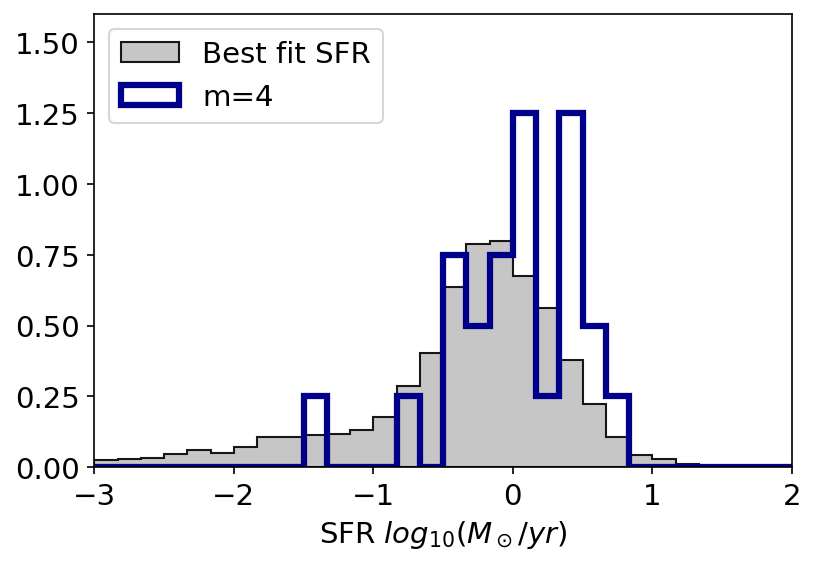}
    \smallskip
        \includegraphics[width=.32\textwidth]{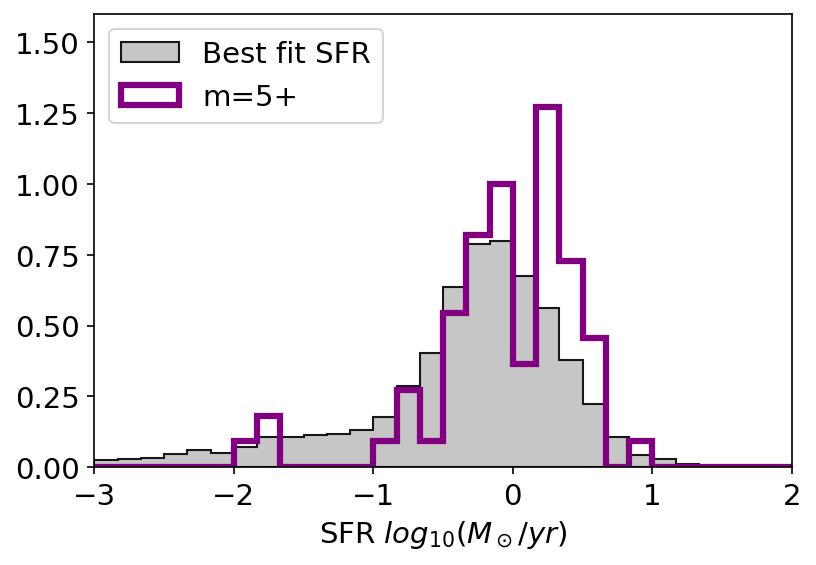}
    
    \caption{SFR histograms for each of the subsamples selected from the limited GAMA-KiDS Galaxy Zoo sample. The gray filled histogram shows the distributions of the entire limited data set, while the colored outlines show the distribution for the individual spiral arm number subsamples.}
    
    \label{pics:sfrfig}
\end{figure*}

    % SFR summary 
\begin{figure}
   \centering
    \includegraphics[width=0.49\textwidth]{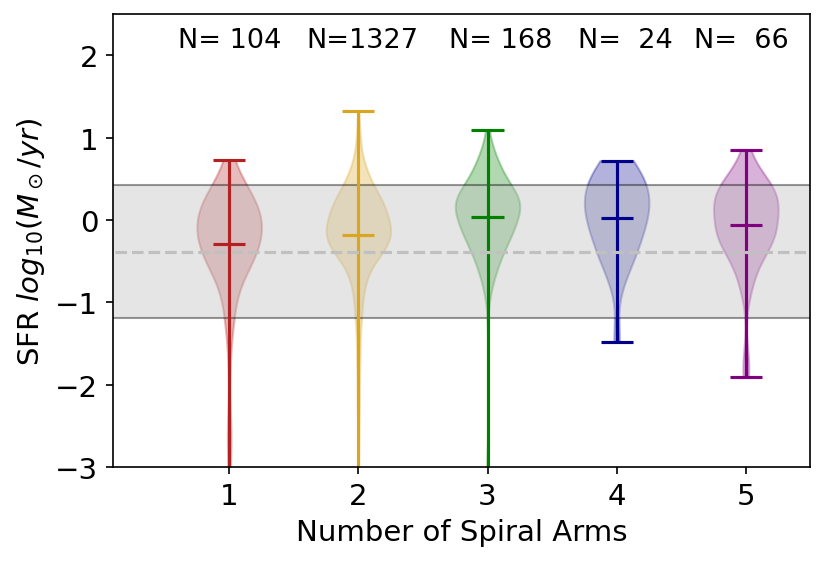}
    \caption{SFR distribution densities for each of the spiral arm subsamples. The shaded regions are equivalent to the definitions in Figure \ref{fig:stmsummary}.}
    \label{fig:sfrsummary}
\end{figure}

    % SSFR histograms
\begin{figure*}
    \centering
        \includegraphics[width=.32\textwidth]{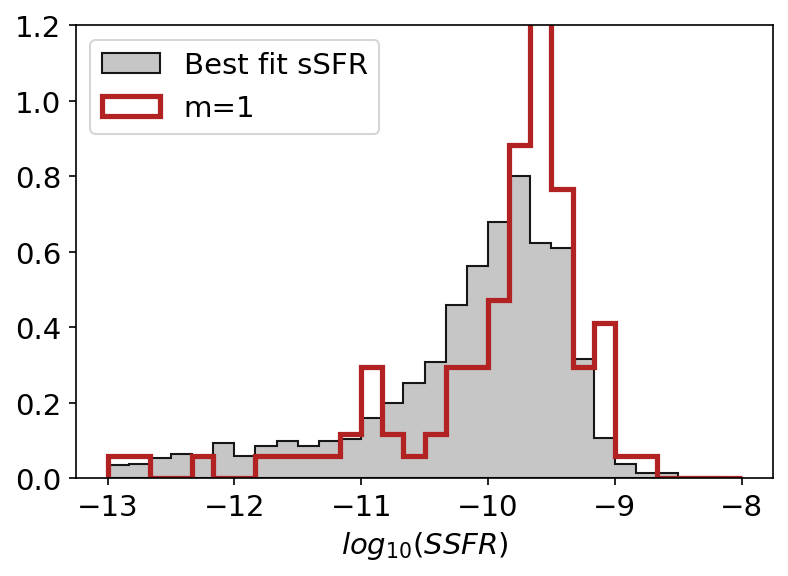}\quad
        \includegraphics[width=.32\textwidth]{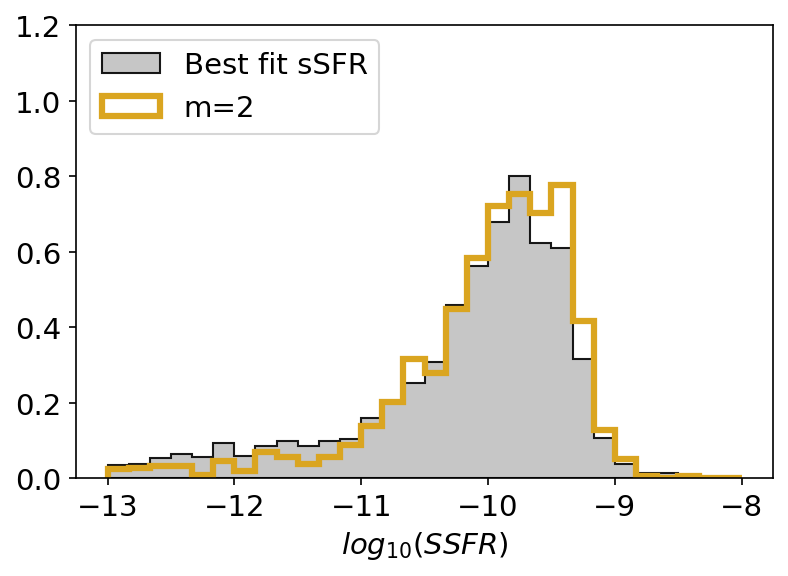}
    \smallskip
        \includegraphics[width=.32\textwidth]{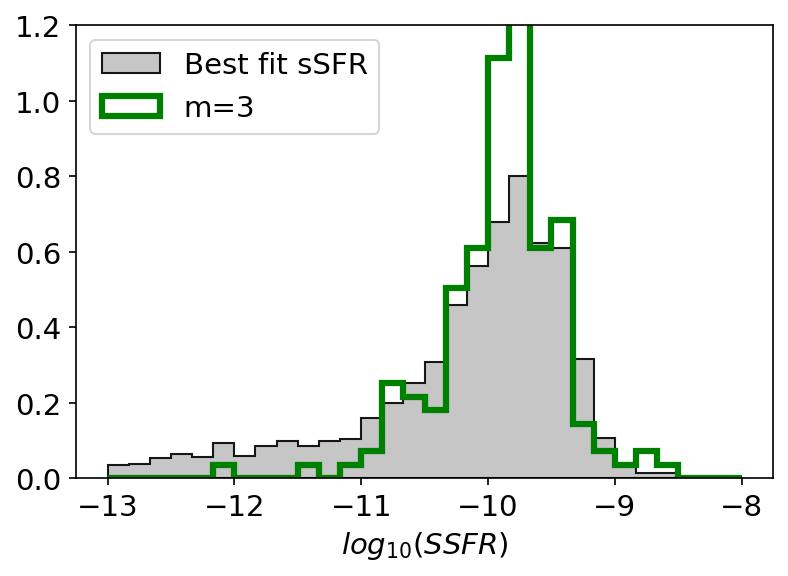}\quad
        \includegraphics[width=.32\textwidth]{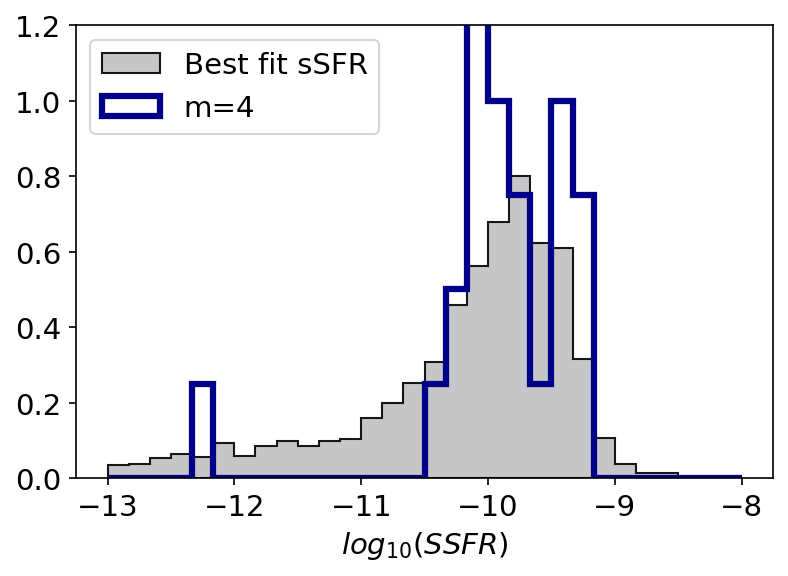}
    \smallskip
        \includegraphics[width=.32\textwidth]{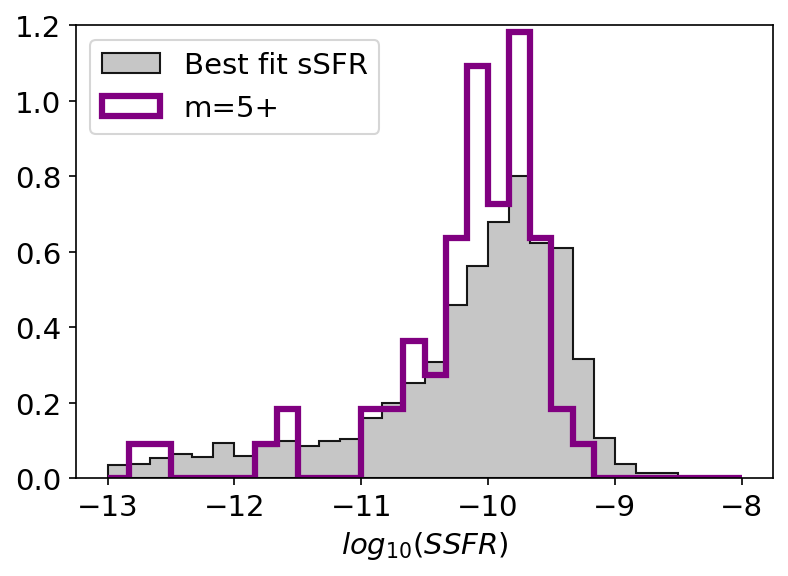}
    
    \caption{sSFR histograms for each of the subsamples selected from the limited GAMA-KiDS Galaxy Zoo sample. The gray filled histogram shows the distributions of the entire limited data set, while the colored outlines show the distribution for the individual spiral arm number subsamples.}
    \label{pics:specificfig}
\end{figure*}  

    % sSFR summary
\begin{figure}
   \centering
    \includegraphics[width=0.49\textwidth]{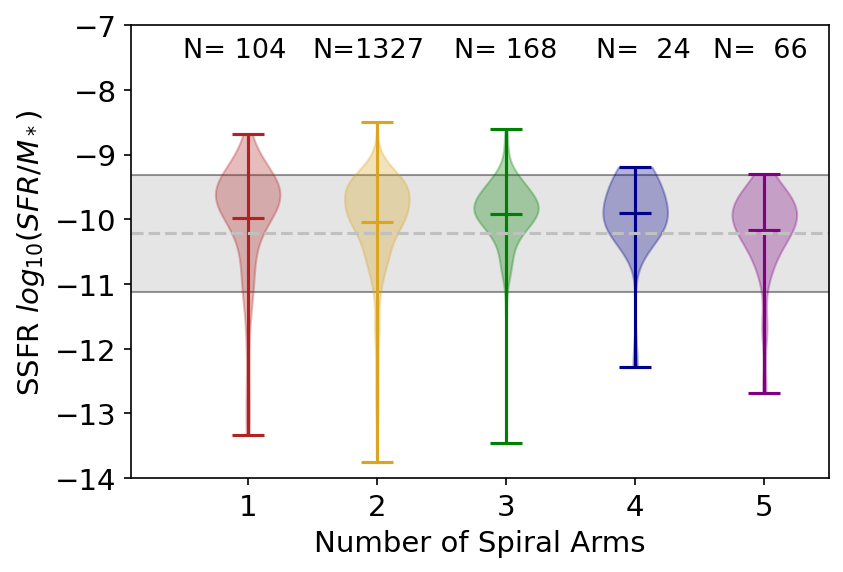}
    \caption{sSFR distribution densities for each of the spiral arm subsamples. The shaded regions are equivalent to the definitions in Figure \ref{fig:stmsummary}.}
    \label{fig:SSFRsummary}
\end{figure}

\subsection{Stellar Mass} \label{results: stellar mass}
Figure \ref{pics:stellarmassfig} shows the histograms resulting from the process described in sections \ref{sample selection} and \ref{subsample}, and from the K-S test described above. 

The m=1, m=3, and m=5 subsamples show visual differences in their stellar mass distributions. The m=5 subsample, though limited by a small number of galaxies, shows a notable shift towards higher stellar masses. Likewise the m=3 subsample tends towards higher masses as well, with the peak falling just below 10.0 for stellar mass, versus the peak at 9.5 for the limited sample. The m=1 subsample shows a tendency to lower stellar masses. 

These are reflected in the K-S statistic in Table \ref{tab:statstable}, with the m=5 subsample having the greatest difference from the limited sample. The m=4 sample has the second highest statistic value, but with the lowest number of galaxies and p-value of .08 on that statistic, we do not consider it as significant. The m=1, 3, and 5 values are significant in their A-D test as well, with slightly higher significance for m=1.

Figure \ref{fig:stmsummary} shows the distributions in the colored violin plots, with the gray band indicating the median and $\pm$ 1 standard deviation for the limited sample. This also reflects the shift in distribution, with m=1 having both a lower median than the limited sample and showing a greater quantity of galaxies at lower stellar masses. Likewise, the m=3 subsample is shifted to slightly higher stellar masses, and m=5 visibly higher than the median from the limited sample. 

\subsection{Star Formation Rate} \label{results: sfr}

As for stellar mass, Figure \ref{pics:sfrfig} shows the histograms resulting from the above process. The m=3, 4, and 5 distributions show a notable difference in their star formation rate distributions, with each of them having a higher SFR distribution than the limited sample. Visually, the m=3 subsample appears to have the highest distribution for SFR, and the K-S statistic reflects a greater difference from the limited sample than most other subsamples. We find that these three subsamples have a notable difference in their distributions from the limited sample, as reflected in Table \ref{tab:statstable} with their K-S statistics being much higher than the m=1 or m=2 samples. This difference is reflected again in their A-D statistics with high significance for the m=2, 3, 4, and 5+ samples. The agreement between K-S and A-D statistics is due to the mostly Gaussian shape of the distributions in SFR with only a weak tail to lower SFR.

Figure \ref{fig:sfrsummary}, as above, shows the summarized distributions for SFR. This reflects a higher average distribution for m=3, 4, and 5 subsamples. Though the m=1 and 2 subsamples appear to have higher SFR distributions than the limited sample in Figures \ref{pics:sfrfig} and \ref{fig:sfrsummary}, they also have relatively low K-S statistics compared to the other subsamples, showing a higher similarity to the limited sample than m=3, 4, or 5. 

\subsection{Specific Star Formation Rate} \label{results:SSFR}
As in sections \ref{results: stellar mass} and \ref{results: sfr}, Figure \ref{pics:specificfig} shows the histograms for the sSFR values. The distribution for subsamples m=4 and m=5 are shifted to lower sSFRs. The K-S statistics for m=4,and 5 reflect these distribution shifts, but we consider the p-values of their K-S statistics to be less significant. 

Conversely, we see a significant shift in the m=1 population towards higher sSFRs, and the m=3 population's distribution weighted heavily towards -10. Again this is reflected by the K-S statistics and high significance p-values in table \ref{tab:statstable}, where the m=1 subsample has a greater difference in the limited sample, and m=3 fitting the limited sample quite well. The A-D tests confirm the significance of different distributions depending on the number of spiral arms and add a significant difference for the m=5+ distribution (5\% critical value exceeded by the A-D test). This lends high confidence to the conclusion that sSFR and arm number are strongly correlated.

As above, Figure \ref{fig:SSFRsummary} shows the summarized distributions for sSFR. We see that m=1 has a  higher than average sSFR compared to the other samples, and that the distribution of m=3 is more concentrated into one peak region. 

%%%% DISCUSSION %%%%
\section{Discussion} \label{discussion}

In section \ref{sample selection}, we detail how the spiral arm number subsamples are defined. In categorizing them based on the Galaxy Zoo votes (question T06, shown in Figure \ref{fig:GZ4}) we are treating spiral arms as integers. However, this does not take into account whether a spiral galaxy has well-defined arms; flocculent spiral galaxies with poorly-defined or discontinuous arms can not be well classified with an integer number of arms. The voting pattern does reflect this somewhat, as the classifications came from real people voting, and so any galaxies with poorly-defined arms would be best categorized through the majority vote. So, a galaxy is classified as accurately as it can be into an integer number of spiral arms. From question T03 (Figure \ref{fig:GZ4}), we do know that these are all spiral galaxies. %We do know, from question T03 (Figure \ref{fig:GZ4}) that these are all spiral galaxies. 

In Figures \ref{pics:stellarmassfig}, \ref{pics:sfrfig}, and \ref{pics:specificfig} we can see that the low number of galaxies in each subsample does leave the m=4 and m=5+ distribution lacking in statistical weight for the K-S test results compared to, for example, the m=2 subsample. \cite{Hart17} used the optical-WISE SED inferred stellar masses and separately estimated star formation rates from either FUV flux or 22$\mu$m flux. The improvement in our data is the use of a self-consistent SED to determine both from 21 filters spanning ultraviolet through sub-mm \citep{Wright16,Driver16c}. Additionally, the A-D test results lend more statistical significance to the m=5+ distribution in particular, while giving a more significant result (to 0.1\%) for sSFR for m=1, 2, and 3. Because of the small sample size of the m=4 subsample, the A-D test (which is more sensitive to the tails of the distribution) does not show a higher significance than the m=4 sSFR result.

% this is reiterated in the last paragraph in this section much more succinctly: %The small sample size for m=4 and m=5+ does affect the distributions shown; in Figures \ref{fig:sfrsummary} and \ref{fig:SSFRsummary} we see that those subsamples have a comparatively small range of values. The overall shape of the distributions with low numbers seem analogous to, but different than, those with a greater number of galaxies. Notably the m=1 and m=5+ subsamples are close to each other in sample size (N=104 and N=66, respectively) but Figure \ref{pics:specificfig} shows that many-armed spirals tend toward lower SSFRs than the limited sample, and one-armed spirals tend toward higher SSFRs. Given the statistical results in Table \ref{tab:statstable} (as discussed in section \ref{results:SSFR}) we find this to be a convincing result. 

Overall, we find that spiral galaxies are less efficient at forming stars if they have more spiral arms. The m=1 subsample has a much lower stellar mass on average, but a higher than average distribution for sSFR (see Figures \ref{pics:stellarmassfig} and \ref{pics:specificfig}). This is supported by the findings of \cite{Hart17}, who noted that two armed spiral galaxies are more gas deficient than other galaxies, and so are more efficient at converting gas to stars. 

Galaxies with stronger bars have fewer but stronger arms \citep{Yu20}, and arm strength has been found to correlate well with SFR as a function of stellar mass \citep{Yu21}. Given our results, it is unclear if the causation is more arms leads to weaker arms which in turn leads to lower sSFR. Alternatively, it is possible that the perceived change in sSFR is caused by a subtle bias in the MAGPHYS SED results (Section \ref{magphys}), because the arm patterns rearrange the dusty interstellar medium (ISM) in the disc, skewing SED measurements of star formation. Arms are more opaque than the disc \citep{kw00b,Holwerda05b}, and therefore better at hiding directly measured star formation. The many-armed spirals with low sSFR might simply be hiding their directly measurable star formation instead of having lower rates overall. However, \cite{Hart17} found that two-armed spirals have more mid-infrared (MIR) dust emission, indicating that a greater proportion of new stars in two-armed spirals are in heavily obscured region and the MAGPHYS SED result is based on balancing the missing ultraviolet light with the observed heated dust emission. Given this, it seems unlikely that low sSFRs in many-armed spirals are caused by a higher obscuration fraction of new stars.

%Alternatively, higher star formation rates will emphasize the spiral pattern in the disc. 
Higher star formation for a given mass will likely highlight the spiral structure in these disks as the site of recent star formation. \cite{Hart17} note that the mean of their distribution shifts with only 0.05 dex with each additional spiral arm. We point to Figure \ref{fig:SSFRsummary} to show that the mode of the distribution is a better indication of the change with the number of arms. Between the shift in the distribution of sSFR values and the much improved star formation and stellar mass accuracy thanks to a consistent SED treatment rather than single-flux based estimates, we find the trend in lowering sSFR with number of spiral arms convincing.

\section{Conclusions} \label{conclusions}

In this paper, we examined the connection of spiral arm number with stellar mass, star formation rate (SFR), and specific star formation rate (sSFR). Using the data from GAMA DR3 and the morphological classifications from Galaxy Zoo GAMA-KiDS, we compared subsamples consisting of galaxies with 1, 2, 3, 4, or 5+ spiral arms. Overall, we find the following: 
\begin{enumerate}
    \item Galaxies with more spiral arms tend towards higher stellar masses (Figure \ref{fig:stmsummary}) and higher star formation rates (Figure \ref{fig:sfrsummary}).
    \item Galaxies with more spiral arms tend towards lower specific star formation rates (Figures \ref{pics:specificfig} and \ref{fig:SSFRsummary}, Table \ref{tab:statstable}).
    \item The single arm (m=1) subsample tends to have lower stellar mass and higher specific star formation than both the full sample and any other subsample.
\end{enumerate}
A different, non-integer classification of the number of spiral arms, allowing for the voting tally to assign fractions of spiral arms to galaxies may reflect the reality of these galaxies better. Additionally, changing the limited sample to include only galaxies with a sufficient number of votes to ensure accuracy in arm classification (as opposed to percentages of votes in favor of spiral arm pattern) may yield a higher sample size with stronger statistical significance. 

The Rubin Observatory and future iterations of the Galaxy Zoo are expected to improve the statistics of spiral arm numbers on galaxies in the nearby Universe. Equally important however, are good stellar mass and star-formation estimates from SED models for similar comparisons as this work and in \cite{Hart17a}. 

The Euclid and Roman space telescopes will a wealth of morphological data on higher-redshift spiral galaxies. These will allow for a direct comparison of the evolution of spiral structure.

\section*{Acknowledgements}
% NASA-KY funding. 
% people that gave us $$
% NASA KY Space Grant 2020-2024 (GF, REU, TP, RIA, MG, EMG): 
The material is supported by NASA Kentucky award No: 80NSSC20M0047 (NASA-REU to L. Haberzettl and R. Porter-Temple).
BWH is supported by an Enhanced Mini-Grant (EMG). The material is based upon work supported by NASA Kentucky under NASA award No: 80NSSC20M0047.

This research made use of Astropy, a community-developed core Python package for Astronomy \citep{Astropy-Collaboration13,Astropy-Collaboration18}. 

\section{Data Availability}
The data for this project is available from the GAMA DR3 website \url{http://www.gama-survey.org/dr3/schema/table.php?id=82}.

%%%%%%%%%%%%%%%%%%%%%%%%%%%%%%%%%%%%%%%%%%%%%%%%%%


%\bibliographystyle{mnras}
%\bibliography{Bibliography} % if your bibtex file is called example.bib

\begin{thebibliography}{}
\makeatletter
\relax
\def\mn@urlcharsother{\let\do\@makeother \do\$\do\&\do\#\do\^\do\_\do\%\do\~}
\def\mn@doi{\begingroup\mn@urlcharsother \@ifnextchar [ {\mn@doi@}
  {\mn@doi@[]}}
\def\mn@doi@[#1]#2{\def\@tempa{#1}\ifx\@tempa\@empty \href
  {http://dx.doi.org/#2} {doi:#2}\else \href {http://dx.doi.org/#2} {#1}\fi
  \endgroup}
\def\mn@eprint#1#2{\mn@eprint@#1:#2::\@nil}
\def\mn@eprint@arXiv#1{\href {http://arxiv.org/abs/#1} {{\tt arXiv:#1}}}
\def\mn@eprint@dblp#1{\href {http://dblp.uni-trier.de/rec/bibtex/#1.xml}
  {dblp:#1}}
\def\mn@eprint@#1:#2:#3:#4\@nil{\def\@tempa {#1}\def\@tempb {#2}\def\@tempc
  {#3}\ifx \@tempc \@empty \let \@tempc \@tempb \let \@tempb \@tempa \fi \ifx
  \@tempb \@empty \def\@tempb {arXiv}\fi \@ifundefined
  {mn@eprint@\@tempb}{\@tempb:\@tempc}{\expandafter \expandafter \csname
  mn@eprint@\@tempb\endcsname \expandafter{\@tempc}}}

\bibitem[\protect\citeauthoryear{{Astropy Collaboration} et~al.,}{{Astropy
  Collaboration} et~al.}{2013}]{Astropy-Collaboration13}
{Astropy Collaboration} et~al., 2013, \aap, 558, A33

\bibitem[\protect\citeauthoryear{{Astropy Collaboration} et~al.,}{{Astropy
  Collaboration} et~al.}{2018}]{Astropy-Collaboration18}
{Astropy Collaboration} et~al., 2018, \aj, 156, 123

\bibitem[\protect\citeauthoryear{{Athanassoula}, {Romero-G{\'o}mez}, {Bosma}
  \& {Masdemont}}{{Athanassoula} et~al.}{2009}]{Athanassoula09}
{Athanassoula} E.,  {Romero-G{\'o}mez} M.,  {Bosma} A.,   {Masdemont} J.~J.,
  2009, \mnras, 400, 1706

\bibitem[\protect\citeauthoryear{{Baldry} et~al.,}{{Baldry}
  et~al.}{2018}]{Baldry18}
{Baldry} I.~K.,  et~al., 2018, \mnras, 474, 3875

\bibitem[\protect\citeauthoryear{{Chang}, {van der Wel}, {da Cunha}  \&
  {Rix}}{{Chang} et~al.}{2015}]{Chang15}
{Chang} Y.-Y.,  {van der Wel} A.,  {da Cunha} E.,   {Rix} H.-W.,  2015,
  preprint

\bibitem[\protect\citeauthoryear{{D{\'\i}az-Garc{\'\i}a}, {Salo}, {Knapen}  \&
  {Herrera-Endoqui}}{{D{\'\i}az-Garc{\'\i}a} et~al.}{2019}]{Diaz-Garcia19c}
{D{\'\i}az-Garc{\'\i}a} S.,  {Salo} H.,  {Knapen} J.~H.,   {Herrera-Endoqui}
  M.,  2019, \aap, 631, A94

\bibitem[\protect\citeauthoryear{{Domingue}, {Keel}  \& {White}}{{Domingue}
  et~al.}{2000}]{kw00b}
{Domingue} D.~L.,  {Keel} W.~C.,   {White} III R.~E.,  2000, \apj, 545, 171

\bibitem[\protect\citeauthoryear{{Driver} et~al.,}{{Driver}
  et~al.}{2009}]{Driver09}
{Driver} S.~P.,  et~al., 2009, Astronomy and Geophysics, 50, 050000

\bibitem[\protect\citeauthoryear{{Driver} et~al.,}{{Driver}
  et~al.}{2011}]{Driver11}
{Driver} S.~P.,  et~al., 2011, \mnras, 413, 971

\bibitem[\protect\citeauthoryear{{Driver} et~al.,}{{Driver}
  et~al.}{2016}]{Driver16c}
{Driver} S.~P.,  et~al., 2016, \mnras, 455, 3911

\bibitem[\protect\citeauthoryear{{Driver} et~al.,}{{Driver}
  et~al.}{2018}]{Driver18}
{Driver} S.~P.,  et~al., 2018, \mnras, 475, 2891

\bibitem[\protect\citeauthoryear{{Hart}, {Bamford}, {Casteels}, {Kruk},
  {Lintott}  \& {Masters}}{{Hart} et~al.}{2017a}]{Hart17}
{Hart} R.~E.,  {Bamford} S.~P.,  {Casteels} K. R.~V.,  {Kruk} S.~J.,  {Lintott}
  C.~J.,   {Masters} K.~L.,  2017a, \mnras, 468, 1850

\bibitem[\protect\citeauthoryear{{Hart} et~al.,}{{Hart}
  et~al.}{2017b}]{Hart17a}
{Hart} R.~E.,  et~al., 2017b, \mnras, 472, 2263

\bibitem[\protect\citeauthoryear{{Hart}, {Bamford}, {Keel}, {Kruk}, {Masters},
  {Simmons}  \& {Smethurst}}{{Hart} et~al.}{2018}]{Hart18}
{Hart} R.~E.,  {Bamford} S.~P.,  {Keel} W.~C.,  {Kruk} S.~J.,  {Masters} K.~L.,
   {Simmons} B.~D.,   {Smethurst} R.~J.,  2018, \mnras, 478, 932

\bibitem[\protect\citeauthoryear{{Holwerda}, {Gonz\'alez}, {Allen}  \& {van der
  Kruit}}{{Holwerda} et~al.}{2005}]{Holwerda05b}
{Holwerda} B.~W.,  {Gonz\'alez} R.~A.,  {Allen} R.~J.,   {van der Kruit} P.~C.,
   2005, \aj, 129, 1396

\bibitem[\protect\citeauthoryear{{Holwerda}, {Fraine}, {Mouawad}  \&
  {Bridge}}{{Holwerda} et~al.}{2019a}]{Holwerda19b}
{Holwerda} B.~W.,  {Fraine} J.,  {Mouawad} N.,   {Bridge} J.~S.,  2019a, \pasp,
  131, 114504

\bibitem[\protect\citeauthoryear{{Holwerda} et~al.,}{{Holwerda}
  et~al.}{2019b}]{Holwerda19}
{Holwerda} B.~W.,  et~al., 2019b, \aj, 158, 103

\bibitem[\protect\citeauthoryear{{Kendall}, {Clarke}  \& {Kennicutt}}{{Kendall}
  et~al.}{2014}]{Kendall14}
{Kendall} S.,  {Clarke} C.,   {Kennicutt} R.~C.,  2014, preprint

\bibitem[\protect\citeauthoryear{{Knabel} et~al.,}{{Knabel}
  et~al.}{2020}]{Knabel20}
{Knabel} S.,  et~al., 2020, \aj, 160, 223

\bibitem[\protect\citeauthoryear{{Kuijken} et~al.,}{{Kuijken}
  et~al.}{2019}]{Kuijken19}
{Kuijken} K.,  et~al., 2019, \aap, 625, A2

\bibitem[\protect\citeauthoryear{{Lingard} et~al.,}{{Lingard}
  et~al.}{2021}]{Lingard21}
{Lingard} T.,  et~al., 2021, \mnras, 504, 3364

\bibitem[\protect\citeauthoryear{{Lintott} et~al.,}{{Lintott}
  et~al.}{2008}]{Lintott08}
{Lintott} C.~J.,  et~al., 2008, \mnras, 389, 1179

\bibitem[\protect\citeauthoryear{{Liske} et~al.,}{{Liske}
  et~al.}{2015}]{Liske15}
{Liske} J.,  et~al., 2015, \mnras, 452, 2087

\bibitem[\protect\citeauthoryear{{Martin} et~al.,}{{Martin}
  et~al.}{2005}]{Martin05}
{Martin} D.~C.,  et~al., 2005, \apjl, 619, L1

\bibitem[\protect\citeauthoryear{{Masters} et~al.,}{{Masters}
  et~al.}{2021}]{Masters21}
{Masters} K.~L.,  et~al., 2021, \mnras, 507, 3923

\bibitem[\protect\citeauthoryear{{Miller}, {Kennefick}, {Kennefick}, {Shameer
  Abdeen}, {Monson}, {Eufrasio}, {Shields}  \& {Davis}}{{Miller}
  et~al.}{2019}]{Miller19a}
{Miller} R.,  {Kennefick} D.,  {Kennefick} J.,  {Shameer Abdeen} M.,  {Monson}
  E.,  {Eufrasio} R.~T.,  {Shields} D.~W.,   {Davis} B.~L.,  2019, \apj, 874,
  177

\bibitem[\protect\citeauthoryear{{Pringle} \& {Dobbs}}{{Pringle} \&
  {Dobbs}}{2019}]{Pringle19a}
{Pringle} J.~E.,  {Dobbs} C.~L.,  2019, \mnras, 490, 1470

\bibitem[\protect\citeauthoryear{{Savchenko}, {Marchuk}, {Mosenkov}  \&
  {Grishunin}}{{Savchenko} et~al.}{2020}]{Savchenko20}
{Savchenko} S.,  {Marchuk} A.,  {Mosenkov} A.,   {Grishunin} K.,  2020, \mnras,
  493, 390

\bibitem[\protect\citeauthoryear{{Seigar} \& {James}}{{Seigar} \&
  {James}}{1998}]{Seigar98}
{Seigar} M.~S.,  {James} P.~A.,  1998, \mnras, 299, 672

\bibitem[\protect\citeauthoryear{{Willett} et~al.,}{{Willett}
  et~al.}{2013}]{Willett13}
{Willett} K.~W.,  et~al., 2013, \mnras, 435, 2835

\bibitem[\protect\citeauthoryear{{Wright} et~al.,}{{Wright}
  et~al.}{2016}]{Wright16}
{Wright} A.~H.,  et~al., 2016, \mnras, 460, 765

\bibitem[\protect\citeauthoryear{{Wright} et~al.,}{{Wright}
  et~al.}{2017}]{Wright17}
{Wright} A.~H.,  et~al., 2017, \mnras, 470, 283

\bibitem[\protect\citeauthoryear{{Yu} \& {Ho}}{{Yu} \& {Ho}}{2018a}]{Yu18}
{Yu} S.-Y.,  {Ho} L.~C.,  2018a, arXiv e-prints, p. arXiv:1812.06010

\bibitem[\protect\citeauthoryear{{Yu} \& {Ho}}{{Yu} \& {Ho}}{2018b}]{Yu18b}
{Yu} S.-Y.,  {Ho} L.~C.,  2018b, \apj, 869, 29

\bibitem[\protect\citeauthoryear{{Yu}, {Ho}, {Barth}  \& {Li}}{{Yu}
  et~al.}{2018}]{Yu18a}
{Yu} S.-Y.,  {Ho} L.~C.,  {Barth} A.~J.,   {Li} Z.-Y.,  2018, \apj, 862, 13

\bibitem[\protect\citeauthoryear{{Yu} et~al.,}{{Yu} et~al.}{2020}]{Yu20}
{Yu} S.,  et~al., 2020, \mnras, 494, 1539

\bibitem[\protect\citeauthoryear{{Yu} et~al.,}{{Yu} et~al.}{2021}]{Yu21}
{Yu} S.,  et~al., 2021, arXiv e-prints, p. arXiv:2103.03888

\bibitem[\protect\citeauthoryear{{da Cunha}, {Charlot}  \& {Elbaz}}{{da Cunha}
  et~al.}{2008}]{da-Cunha08}
{da Cunha} E.,  {Charlot} S.,   {Elbaz} D.,  2008, \mnras, 388, 1595

\bibitem[\protect\citeauthoryear{{de Jong}, {Verdoes Kleijn}, {Kuijken}  \&
  {Valentijn}}{{de Jong} et~al.}{2013}]{de-Jong13}
{de Jong} J.~T.~A.,  {Verdoes Kleijn} G.~A.,  {Kuijken} K.~H.,   {Valentijn}
  E.~A.,  2013, Experimental Astronomy, 35, 25

\bibitem[\protect\citeauthoryear{{de Jong} et~al.,}{{de Jong}
  et~al.}{2015}]{de-Jong15}
{de Jong} J.~T.~A.,  et~al., 2015, \aap, 582, A62

\bibitem[\protect\citeauthoryear{{de Jong} et~al.,}{{de Jong}
  et~al.}{2017}]{de-Jong17}
{de Jong} J.~T.~A.,  et~al., 2017, \aap, 604, A134

\makeatother
\end{thebibliography}


%%%%%%%%%%%%%%%%%%%%%%%%%%%%%%%%%%%%%%%%%%%%%%%%%%

%%%%%%%%%%%%%%%%% APPENDICES %%%%%%%%%%%%%%%%%%%%%

% \appendix

% \section{Some extra material}

% If you want to present additional material which would interrupt the flow of the main paper,
% it can be placed in an Appendix which appears after the list of references.

%%%%%%%%%%%%%%%%%%%%%%%%%%%%%%%%%%%%%%%%%%%%%%%%%%

% Don't change these lines
\bsp	% typesetting comment
\label{lastpage}
\end{document}